\documentstyle[aps,epsf,bookmath,multicol]{revtex}
\begin{document}

\vspace*{-2cm}\hspace*{11cm}{\it submitted to Nature}
\vspace*{2cm}

\centerline{\bf Discovery of the Acoustic Faraday Effect in Superfluid
$^3$He-B}
\centerline{Y. Lee, T.M. Haard, W.P. Halperin and J.A. Sauls}
\centerline{\small\it Department of Physics and Astronomy,
	 Northwestern University, Evanston, IL 60208, USA}
\date{\today}

\bigskip
{\bf
Acoustic waves provide a powerful tool for studying the structure
of matter. The speed, attenuation
and dispersion of acoustic waves give useful details of the
molecular forces  and microscopic mechanisms for absorption and
scattering of acoustic energy. In solids both compressional and
shear waves occur, so-called longitudinal and transverse
sound.  However, normal liquids do not support shear forces and
consequently transverse waves do not propagate in liquids with one
notable exception. In 1957 Landau predicted \cite{lan57b} that
the quantum liquid  phase of $^3$He might exhibit transverse
sound at sufficiently low temperatures where the restoring
forces for shear waves are supplied by the collective action of the
particles in the fluid. Shear waves in liquid
$^3$He involve displacements of the  fluid transverse
to the direction of propagation. The displacement defines the polarization
direction of the wave similar to electromagnetic waves. We have observed
rotation of the polarization of transverse sound waves in superfluid
$^3$He-B in a magnetic field. This magneto-acoustic effect is
the direct analogue to the magneto-optical effect discovered by
Michael Faraday in 1845, where
the polarization of an electromagnetic wave is rotated by a
magnetic field along the propagation direction.}

Superfluidity in $^3$He results from the binding of the
$^3$He particles with nuclear spin $s=1/2$ into molecules 
called ``Cooper pairs'' with binding energy, 
$2\Delta$.~\cite{leg75,vol90,and75,whe75} The pairs
undergo a type of Bose-Einstein condensation having a
close analogy to the Bardeen-Cooper-Schrieffer
condensation~\cite{bar57}  phenomenon associated with
superconductivity in metals. One important difference
is that the pairs that form the condensate in $^3$He have
total spin $S=1$ and an orbital wave function with relative angular
momentum $L=1$ ($p$-$wave$).  This is in contrast to superconductors which
are formed with Cooper pairs of electrons having $S=0$ and 
$L=0$ ($s$-$wave$) or, as is the
case of high temperature superconductors, $S=0$ and $L=2$ ($d$-$wave$). 
In superfluid $^3$He, the spin and orbital angular momentum
vectors are locked at a fixed angle to one another. This is called {\it
broken relative spin-orbit symmetry}.\cite{leg75,vol90}
The equilibrium superfluid state is described as a
condenstae of Cooper pairs with a total angular momentum,
$J=L\oplus S=0$. In addition, the Cooper pairs can be resonantly excited by
sound waves to quantum states with total angular momentum
$J=2$.~\cite{mak74} This is reminiscent of diatomic molecules which have
similar excited states. The above description applies to the B-phase of
superfluid $^3$He, the most stable phase at low pressure. 
The acoustic Faraday effect occurs in $^3$He-B as a consequence of
spontaneously broken relative spin-orbit symmetry.\cite{moo93} An applied
 field magnetically polarizes the spins of the Cooper pairs which, through
coupling to their orbital motion, rotates the polarization of transverse sound.
The rotational excitations of Cooper pairs are essential~\cite{moo93} to
our observation of the propagation of transverse acoustic waves in $^3$He-B
since they significantly increase the sound velocity making the sound mode much
easier to detect; the closer the
sound energy is to the energy of the Cooper pair excited state, the
stronger is this effect.  Furthermore the Cooper pair excited states have a 
linear Zeeman splitting with magnetic field.\cite{ave80,sch81} 
Of the five ($2J+1$) Zeeman sub-states there is one,
$m_J=+1$, which couples to right circularly polarized transverse sound, and
a second, $m_J=-1$, couples to left circularly polarized sound. Thus the speeds of
these two transverse waves are different in a magnetic field. We call this
property acoustic birefringence. It leads to the acoustic Faraday effect where the
magnetic field rotates the polarization direction of linearly polarized sound. Our
measurements show that the rotation angle can be as large as
$1.4\times10^7$~deg/cm-Tesla, much larger than the ususal magneto-optical
Faraday effect.~\cite{ben65}

Excitation and detection of transverse sound is provided by a high Q
($\approx 3000$)
AC-cut, quartz transducer with a fundamental resonance frequency of 12~MHz.
It generates and detects shear waves with a specific linear polarization. The
detection method is based on measurement of the electrical
impedance of the transducer using a frequency-modulated cw-bridge
spectrometer.~\cite{lee96} All measurements were performed
at 82.26~MHz, the $7^{th}$ harmonic of the transducer, with frequency
modulation at 400~Hz and an amplitude of 3~kHz. The electrical impedance 
of the transducer is a direct measure of the acoustic
impedance of the surrounding liquid $^3$He in the
acoustic cavity that is shown in Fig.~1.
Linearly polarized waves are excited by the transducer, 
reflected from the opposite surface of the acoustic cell, and detected
by the same transducer.  Under conditions of high attenuation there is 
no reflected wave, and the acoustic response is determined by the bulk
acoustic impedance, $Z_a=\rho\omega/q$ where $\rho$ is
the density of the liquid, $\omega$ is the sound frequency, $q=k+i\alpha$
is the complex wave number, $\alpha$ is the attenuation, and $2\pi/k$ is the
wavelength. A change in either the attenuation or the phase velocity,
$C_{\phi}=\omega/k$, produces a change in the impedance, $Z_a$.
On cooling into the superfluid the acoustic response shown in Fig.~2 varies
smoothly with temperature in this highly attenuating region. If the attenuation
is low, there is interference between the source and reflected waves which
modulates the local acoustic impedance detected by the transducer. 
Consequently the acoustic response oscillates as the phase velocity changes
with temperature. The oscillations in Fig.~2 at low temperatures correspond to
interference between outgoing and reflected waves, and so they indicate the
existence of some form of propagating wave. Each period of the oscillations
corresponds to a change in velocity sufficient to increase, or decrease, by
unity the number of half  wavelengths in the cavity. The amplitude of the
oscillations increases as the temperature is reduced indicating that 
attenuation of the sound mode decreases with decreasing temperature.

The features labeled $A$ and $B$ in Fig.~2 are identified with known physical
processes for sound absorption in superfluid $^3$He-B.~\cite{vol90,wol78,mck90}
Feature $A$ corresponds to onset of the dissociation of Cooper pairs by sound
where $\hbar\omega=2\Delta(T_A)$. In the temperature range between
$T_A$ and the superfluid transition temperature, $T_c$, the attenuation of
the liquid is extremely high owing to this mechanism.  The point
$B$ corresponds to resonant absorption of sound at $\hbar\omega=
1.5\Delta(T)$ by the excited Cooper pairs with angular momentum 
$J=2$.~\cite{moo93} Transverse sound is extinguished below this temperature.  
Early attempts to observe transverse sound in the normal  phase of
$^3$He were inconlusive,~\cite{roa76a,flo76,flo78}
and furthermore, it was originally expected that the transverse mode would be
suppressed in the superfluid phase.~\cite{com76,mak77}
More recent theoretical work~\cite{moo93} clarified the role
of the Cooper pair excitations showing that they increase the transverse
sound speed which results in a more robust propagating transverse
acoustic wave at low temperatures in superfluid $^3$He-B. The first experimental
evidence for this can be found in the acoustic impedance measurements of Kalbfeld,
Kucera, and Ketterson.~\cite{kal93}

The proof that the impedance oscillations correspond to a propagating 
{\it transverse} sound mode is given in Fig.~3. In  Fig.~3{\bf a}
we show data sets at a pressure of $4.42\, \rm{bar}$ in magnetic fields of
$52\,\rm{G}$, $101\,\rm{G}$ and $152\,\rm{G}$. The principal feature is
that the magnetic field modulates the zero field oscillations shown in
Fig.~2. Our detector is only sensitive to linearly polarized transverse
sound having a specific direction. Application of a field of
$52\,\rm{G}$ in the direction of wave propagation suppresses the oscillations
near $T=0.465\,T_c$ that were present in zero field. This corresponds to a
$90^{\circ}$ rotation of the polarization of the first reflected transverse
sound wave making the polarization orthogonal to the detection direction. 
Doubling the magnetic field restores the transverse sound oscillations at 
this temperature. The oscillations are suppressed once again
by tripling the field to $152\,\rm{G}$. Also note that near the points labeled
$90^{\circ}$ and $270^{\circ}$, there are smaller amplitude impedance oscillations
with shorter period than the primary oscillations. These come from interference of
doubly reflected waves within the cavity. We demonstrate this fact with a simple,
but powerful, simulation of the acoustic impedance oscillations shown in  
Fig.~3{\bf b}.

In zero field, superfluid $^3$He-B is non-magnetic and non-birefringent. Linearly 
polarized transverse sound is the superposition of two circularly polarized waves
having the same velocity and attenuation.  Application of a magnetic field gives 
rise to acoustic {\it circular} birefringence through the Zeeman splitting of the
excited states of the Cooper pairs that couple to the transverse sound modes;
thus, right- and left-circularly polarized waves propagate with different speeds,
$C_{\pm}=C_{\phi}\pm\delta C_{\phi}$. For magnetic fields well below 
$1\,\mbox{kG}$ the difference in propagation speeds is linear in the magnetic field,
$\delta C_{\phi}\propto H$. This implies that a linearly polarized wave generated by
the transducer undergoes Faraday rotation of its polarization as it propagates.
Upon reflection from the opposite wall of the cavity the linearly polarized wave
with ${\bf q}\parallel{\bf H}$ reverses direction. The reflected wave propagates 
with the polarization rotating with the {\it same handedness
relative to the direction of the field}, i.e. the rotation of the polarization
accumulates after reflection from a surface. The spatial period for rotation of
the polarization by $360^{\circ}$ is

\begin{equation}\label{Period}
\Lambda=
4\pi{\left(C_{\phi}\over\omega\right)}{\left|C_{\phi}\over\delta
C_{\phi}\right|}
\,.
\end{equation}

\noindent The Faraday effect produces a sinusoidal modulation of the impedance
oscillations as a function of magnetic field with a period that is inversely
proportional to the field, i.e.  $\Lambda\propto 1/H$.
The constant of proportionality in magneto-optics
is called the Verdet constant, $\mbox{V}=2\pi/H\Lambda$.

In Fig.~3{\bf b} we show the result of our numerical calculation of the sound wave
amplitude in the direction detected by the transducer. The oscillations shown in the
figure come from interference between the source wave and multiply reflected waves.
The calculation uses the attenuation and phase velocity measured in zero field.
The Verdet constant is obtained from the measurement at $52\,\mbox{G}$.
The simulation reproduces all the observed features of the
impedance as a function of temperature including
the maximum in the modulation at $T/T_c=0.415\,,\,H=101\,\mbox{G}$
and the minimum at $T/T_c=0.415\,,\,H=152\,\mbox{G}$, which
confirms that the Faraday period is proportional to $1/H$.
The simulation also produces the fine structure oscillations in the impedance
near the points labeled $90^{\circ}$ and $270^{\circ}$.
The fine structure is observed when the polarization rotates by an
odd multiple of $90^{\circ}$ upon a single round trip in the cell.
Then waves that traverse the cell
twice are $180^{\circ}$ out of phase relative to the source wave,
and consequently the period of the impedance oscillations is halved.
The amplitude of the oscillations is substantially reduced
because of attenuation over the longer pathlength.
This structure provides proof that impedance oscillations are modulated
by the Faraday effect for propagating transverse waves.

The impedance data from our experiments were analyzed to obtain
the spatial period for the rotation of the polarization
and were found to be in agreement with the theoretical
prediction~\cite{moo93} for the Faraday rotation period. The
theoretical results for the period can be expressed in the form,

\begin{equation}\label{Lambda-theory}
\Lambda=K{\sqrt{T/T_{+}-1}\over gH}.
\end{equation}

\noindent for fields $H \ll 1\,\mbox{kG}$
and temperatures above and  near the extinction point $B$. The temperature,
$T_{+}$, corresponds to the extinction of transverse sound by resonant
excitation of Cooper pairs with $J=2, m_J=+1$, at a slightly higher
temperature than the $B$ extinction point in zero field as shown in the
inset to Fig.~2 ($e.g.$ at $H=100\,\mbox{G}$, $T_{+}-T_B
\approx 1~{\mu}K$). The magnitude of the Faraday rotation period
depends on accurately known superfluid properties, contained in the
parameter $K$, as
well as one parameter that is not well-established, the Land\'{e} g-factor,
$g$, for the Zeeman splitting of the $J=2$, Cooper pair excited state.

Movshovich,{\it et al.}\cite{mov88} analyzed the splitting of the
$J=2$ multiplet in the absorption spectrum of longitudinal sound to find a
value of $g=0.042$. In that experiment it was not possible to resolve the
splitting except for fields above $2\,\mbox{kG}$. At these high fields
the non-linear field dependence due to the Paschen-Back effect\cite{sch83,shi83}
becomes comparable to the linear Zeeman splitting,\cite{hal90,mov91}
which makes it difficult to determine the Land\'{e} g-factor accurately.
We have analyzed our measurements of the acoustic Faraday effect to
determine the g-factor with high accuracy at low fields, which eliminates 
the complication of the high-field Paschen-Back effect.
We find $g=0.020\pm 0.002$. Our significantly smaller value of the 
Land\'{e} g-factor has the interpretation that there are important 
$L=3$ ($f$-$wave$) pairing correlations in the
superfluid condensate, about $7\%$ of the dominant $p$-$wave$
interactions.\cite{sau82}

\newpage

\begin{figure}
\caption{Acoustic cavity for transverse sound. Our short path-length acoustic
cavity consists of two quartz transducers 6.3~mm in diameter, labeled (a) and (b).
Their ground planes are face-to-face and the spacing between the transducers is
established by two parallel gold-coated tungsten wires (c). One of the cavity walls
is a 12~MHz, AC-cut transducer (b) used for transverse sound excitation and detection.
The other face of the cavity is a 17~MHz, X-cut transducer (a) for excitation
and detection of longitudinal sound, which we used to determine the cavity
length of $31~\mu m$. The acoustic cavity is immersed in liquid $^3$He at
a pressure near 4.5 bar. The experimental cell is
cooled by a copper nuclear demagnetization refrigerator (not shown).
A superconducting solenoid (d) is placed outside of the sample liquid
enclosed within a superconducting shield.
Temperatures were determined by SQUID based
lanthanum-diluted cerium-magnesium-nitrate thermometry.}
\label{Cell}
\end{figure}
\bigskip
\centerline{\epsfysize=0.75\hsize{\epsfbox{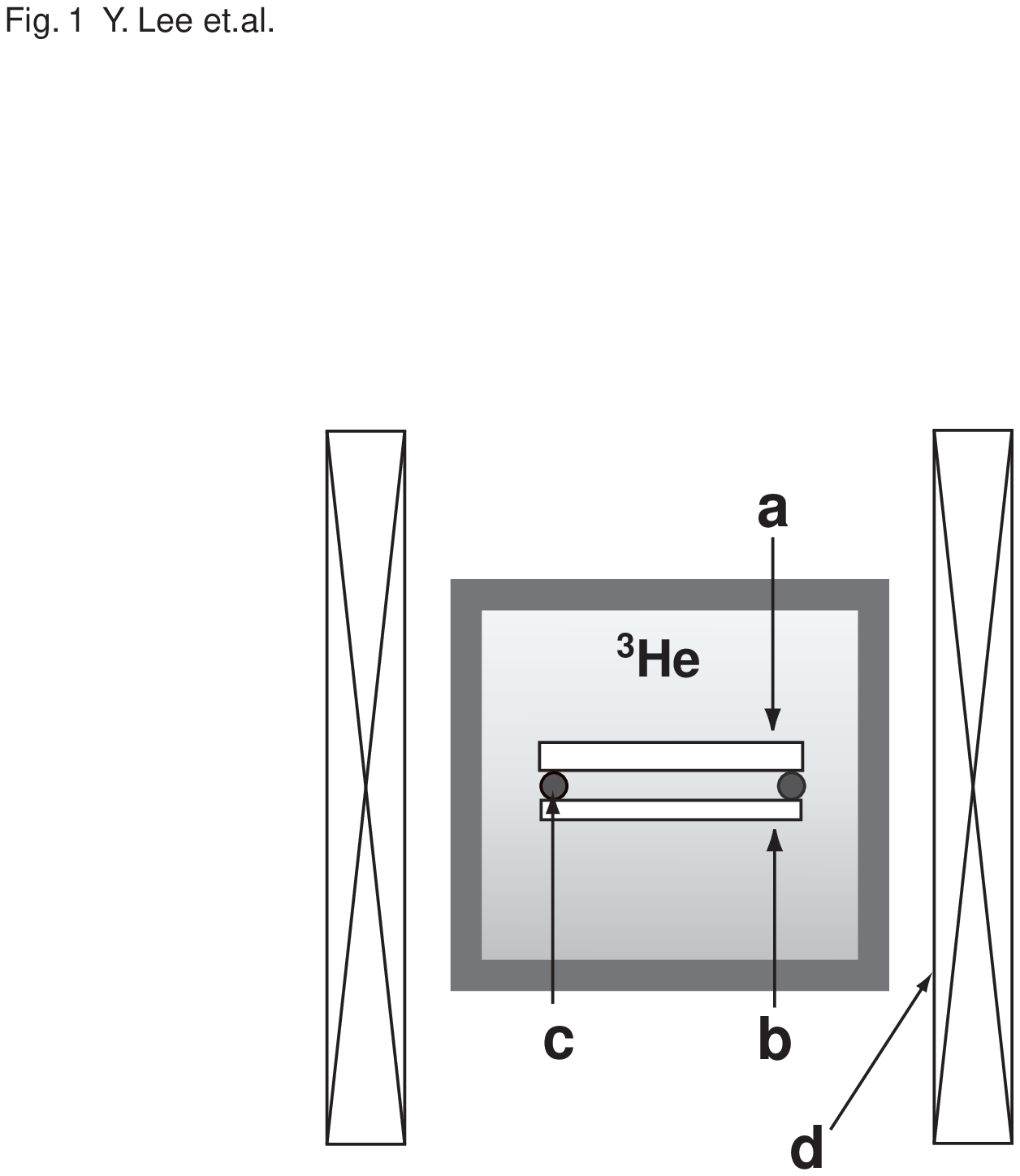}}}

\newpage

\begin{figure}
\caption{Temperature dependence of the acoustic cavity response. Measurements of
the transverse acoustic impedance are shown at a pressure of 4.31~bar in zero
magnetic field. The feature labeled {\it A} corresponds to acoustic pair-breaking,
$\hbar\omega=2\Delta(T_A)$. Impedance oscillations develop for temperatures
$T< 0.6T_c$ and grow in amplitude at lower temperatures as the attenuation of
transverse waves decreases. Near $T\simeq T_B=0.41\,T_c$ the propagating mode is
extinguished by absorption of transverse sound via resonant excitation of Cooper pairs
with total angular momentum $J=2$ at $\hbar\omega=1.5\Delta(T_B)$.
The pair-breaking and resonance conditions of $J=2$ Cooper pair excitations
for fixed frequency are shown in the inset.}
\label{Impedance}
\end{figure}
\bigskip
\centerline{\epsfysize=0.75\hsize{\epsfbox{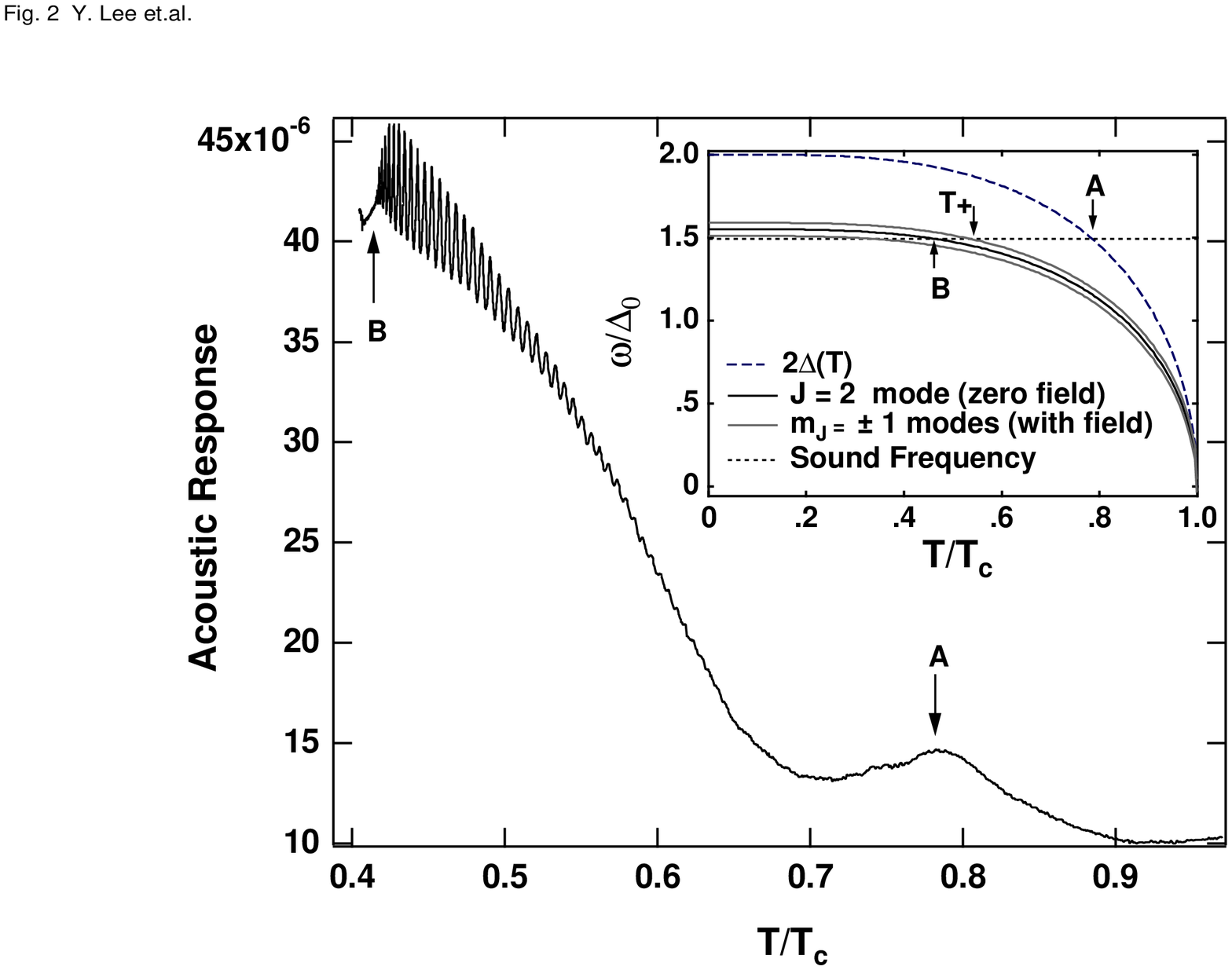}}}

\newpage

\begin{figure}
\caption {Magnetic field dependence of the acoustic cavity response. 
{\bf a}: The sound amplitude measured at 4.42~bar. The angles are indicated 
for rotation of the polarization of transverse sound by 90, 180, and
$270^{\circ}$. {\bf b}: Simulation of the acoustic impedance for the same
path length cell and same temperature, pressure and magnetic fields. The wavelength 
and attenuation of transverse sound in zero field were used as the input data for the
calculation. The Verdet constant for the simulation was determined by the position
of the minimum in the impedance oscillations at $H=52$~G shown in the left panel.}
\label{Faraday}
\end{figure}
\bigskip
\epsfysize=0.75\hsize{\epsfbox{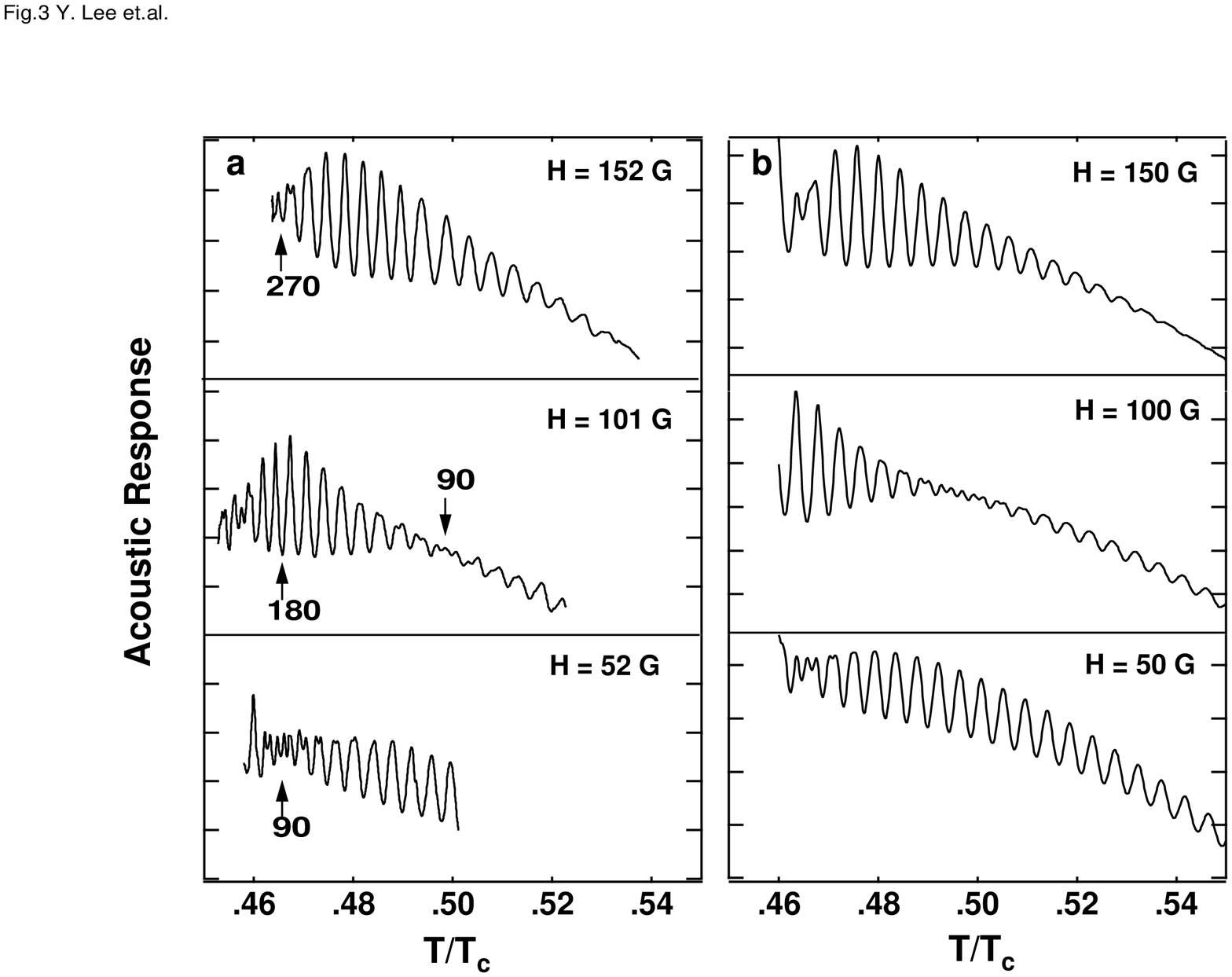}}


\newpage

\begin{thebibliography}{10}

\bibitem{lan57b}
Landau, L.~D.
\newblock Oscillations in a Fermi liquid.
\newblock {\em Sov. Phys. JETP} {\bf 5}, 101-108 (1957).

\bibitem{leg75}
Leggett, A.~J.
\newblock A theoretical interpretation of the new phases of liquid $^3${H}e.
\newblock {\em Rev. Mod. Phys.}, {\bf 47}, 331-414 (1975).

\bibitem{vol90}
Vollhardt, D.~ and W\"olfle, P.
\newblock {\em The {S}uperfluid {P}hases of {$^3${H}e}}.
\newblock (Taylor \& Francis, New York, 1990).

\bibitem{and75}
Anderson, P.~W.  and  Brinkman, W.~F.
\newblock Theory of anisotropic superfluidity in $^3${H}e.
\newblock in {\em {H}elium Liquids}. ed. J.~G. M.~Armitage and I.~E.
Farquhar, p.~315
(Academic Press, New York, 1975).

\bibitem{whe75}
Wheatley, J.~C.
\newblock Experimental properties of superfluid $^3${H}e.
\newblock {\em Rev. Mod. Phys.} {\bf 47}, 415-470 (1975).

\bibitem{bar57}
Bardeen, J.~, Cooper,  L.~N.,  and Schrieffer, R.
\newblock {T}heory of {S}uperconductivity.
\newblock {\em Phys. Rev.} {\bf 108}, 1175-1204 (1957).

\bibitem{mak74}
Maki, K.
\newblock Propagation of zero sound in the {B}alian-{W}erthamer state.
\newblock {\em J. Low Temp. Phys.} {\bf 16}, 465-477 (1974).

\bibitem{moo93}
Moores, G.~F., and  Sauls, J.~A.
\newblock Transverse {W}aves in {S}uperfluid {\heb}.
\newblock {\em J. Low Temp. Phys.} {\bf 91},13-37 (1993).

\bibitem{ave80}
Avenel, O., Varoquaux, E., and Ebisawa, H.
\newblock Field splitting of the new attenuation peak in $^3${H}e-B.
\newblock {\em Phys. Rev. Lett.},  {\bf 45}, 1952-1955  (1980).

\bibitem{sch81}
Schopohl, N.~ and Tewordt, L.
\newblock Land\'{e} factors of collective mode multiplets in $^3${H}e-B and
coupling
strengths to sound waves.
\newblock {\em J. Low Temp. Phys.} {\bf 45}, 67-90 (1981).

\bibitem{ben65}
Bennett, H.~S.  and Stern, E.~A.
\newblock Faraday and Kerr effects in ferromagnetics.
\newblock {\em Phys. Rev.} {\bf 97}, A448-461 (1965).

\bibitem{lee96}
Lee, Y., Hamot, P.~J., Meisel, M.~W., Sprague, D.~T., Haard, T.~M., Kycia,
J.~B.,
Rand, M.~R., and W.P. Halperin
\newblock High frequency acoustic measurements in liquid $^3${H}e near the
  transition temperature.
\newblock {\em J. Low Temp. Phys.} {\bf 103}, 265-272 (1996).

\bibitem{wol78}
W\"olfle, P.~ and Einzel, D.
\newblock Transport and relaxation properties of superfluid $^3${H}e. II.
\newblock {\em J. Low Temp. Phys.} {\bf 32}, 39-56 (1978).

\bibitem{mck90}
Mc{K}enzie, R.~H.  and  Sauls, J.~A.
\newblock {C}ollective {M}odes and {N}onlinear {A}coustics in {S}uperfluid
  {{$^3{H}e$-{B}}}.
\newblock in {\em {H}elium Three}. ed. W.~P.~Halperin and L.~P. Pitaevskii,
p.~255
(Elsevier Science Publishers, Amsterdam, 1990).

\bibitem{roa76a}
Roach, P.~R.  and Ketterson, J.~B.
\newblock Observation of transverse zero sound in normal $^3${H}e.
\newblock {\em Phys. Rev. Lett.},  {\bf 36}, 736-740  (1976).

\bibitem{flo76}
Flowers, E.~G., Richardson, R.~W., and Williamson, S.~J.
\newblock Transverse zero sound in normal $^3${H}e.
\newblock {\em Phys. Rev. Lett.} {\bf 37}, 309-311 (1976).

\bibitem{flo78}
Flowers, E.~G.  and Richardson, R.~W.
\newblock Transverse acoustic impedance of normal $^3${H}e.
\newblock {\em Phys. Rev.} {\bf 17},1238-1248 (1978).

\bibitem{com76}
Combescot, M.~ and Combescot, R.
\newblock Transverse zero sound propagation in superfluid $^3${H}e.
\newblock {\em Phys. Lett.} {\bf 58A},181-182 (1976).

\bibitem{mak77}
Maki, K.,  and Ebisawa, H.
\newblock Transverse zero sound in superfluid $^3${H}e.
\newblock {\em J. Low Temp. Phys.} {\bf 26}, 627-636 (1977).

\bibitem{kal93}
Kalbfeld, S., Kucera, D.~M., and Ketterson, J.~B.
\newblock Observation of an evolving standing-wave pattern involving a
  transverse disturbance in superfluid $^3${H}e.
\newblock {\em Phys. Rev. Lett.} {\bf 71}, 2264-2267  (1993).

\bibitem{mov88}
Movshovich, R., Varoquaux, E.,  Kim, N., and Lee, D.~M.
\newblock Splitting of the squashing collective mode of superfluid $^3${H}e-{B}
  by a magnetic field.
\newblock {\em Phys. Rev. Lett.} {\bf 61}, 1732-1735 (1988).

\bibitem{sch83}
Schopohl, N., Warnke, M., and Tewordt, L.
\newblock Effect of gap distortion on the field splitting of collective
modes in
superfluid $^3${H}e-{B}.
\newblock {\em Phys. Rev. Lett.},  {\bf 50}, 1066-1069  (1983).

\bibitem{shi83}
Shivaram, B.~S., Meisel, M.~W, Sarma, B.~K., Halperin, W.~P., and
Ketterson, J.~B.
\newblock Nonlinear Zeeman shifts in the collective-mode spectrum of
$^3${H}e-{B}.
\newblock {\em Phys. Rev. Lett.},  {\bf 50}, 1070-1072  (1983).

\bibitem{hal90}
Halperin, W.~P.  and Varoquaux, E.
\newblock Order {P}arameter {C}ollective {M}odes in {S}uperfluid $^3${H}e.
\newblock In W.~P. Halperin and L.~P. Pitaevskii, editors, {\em {H}elium
  Three}, p.~353 (Elsevier Science Publishers, Amsterdam, 1990).

\bibitem{mov91}
Movshovich, R., Varoquaux, E.,  Kim, N.,  and Lee, D.~M.
\newblock Fivefold splitting of the squashing mode of superfluid $^3${H}e-{B}
  by a magnetic field.
\newblock {\em Phys. Rev.} {\bf B44}, 332-340 (1991).

\bibitem{sau82}
Sauls, J.~A.  and Serene, J.~W.
\newblock Interaction effects on the {Z}eeman splitting of collective modes in
  superfluid $^3${H}e-{B}.
\newblock {\em Phys. Rev. Lett.} {\bf 49}, 1183-1186 (1982).

\end{thebibliography}
\end{document}